\documentclass[fleqn,usenatbib]{mnras}

\usepackage{newtxtext,newtxmath}

\usepackage[T1]{fontenc}

\DeclareRobustCommand{\VAN}[3]{#2}
\let\VANthebibliography\thebibliography
\def\thebibliography{\DeclareRobustCommand{\VAN}[3]{##3}\VANthebibliography}

\usepackage{graphicx}	
\usepackage{amsmath}	

\usepackage{amssymb}	

\title[Measure Hubble Constant from Fast Radio Bursts]{A Measurement of Hubble Constant Using Cosmographic Approach from Fast Radio Bursts and SNe Ia}

\author[Jiaze Gao et al.]{
	Jiaze Gao$^{1}$\thanks{E-mail: gaojiaze@mail.dlut.edu.cn},
	Zhihuan Zhou$^{1}$,
	Minghui Du$^{2}$,
	Rui Zou$^{3,4}$,
	Jianping Hu$^{5}$,
	Lixin Xu$^{1}$\thanks{Corresponding author. E-mail: lxxu@dlut.edu.cn}
	\\
	$^{1}$Institute of Theoretical Physics, School of Physics,
	Dalian University of Technology, Dalian 116024, People's Republic of China. \\
	$^{2}$Center for Gravitational Wave Experiment, National Microgravity Laboratory, Institute of Mechanics, \\
	\ \ Chinese Academy of Sciences, Beĳing 100190, People's Republic of China.\\
	$^{3}$Department of Physics, Chongqing University, Chongqing 401331, China. \\
	$^{4}$Chongqing Key Laboratory for Strongly Coupled Physics, Chongqing University, Chongqing 401331, China. \\
	$^{5}$School of Astronomy and Space Science, Nanjing University, Nanjing 210093, China\\
}

\date{Accepted XXX. Received YYY; in original form ZZZ}

\pubyear{2015}

\begin{document}
	\label{firstpage}
	\pagerange{\pageref{firstpage}--\pageref{lastpage}}
	\maketitle

\begin{abstract}
	The Hubble constant ${H}_0$ is a crucial parameter in cosmology. However, different cosmic observations have resulted in varying posterior results for ${H}_0$, leading to what is known as the ${H}_0$ tension. In order to address this issue, it is beneficial to use other dataset to constrain ${H}_0$. In this paper, via the cosmographic approach based on the Friedman-Lemaitre-Robertson-Walker (FLRW) metric to the dispersion measure of the intergalactic medium ${\rm{DM}}_{\rm{IGM}}(z)$ of Fast Radio Bursts (FRBs), we obtain the Taylor expansion of $\langle{\rm{DM}}_{\rm{IGM}}(z)\rangle$ in terms redshift $z$. The result for Hubble constant $H_0=65.5^{+6.4}_{-5.4}$ ${\rm{km~s^{-1}~Mpc^{-1}}}$ $(68$$\%$ ${\rm{C.L.}}) $, cosmological deceleration parameter $q_0=-0.50\pm 0.20 $ and the jerk parameter $j_0=-0.1^{+2.0}_{-2.5}$ using uncalibrated Supernova Ia (SNe Ia) Pantheon dataset combined with 18 localized FRBs are obtained. To demonstrate the impact of parameter degeneracies on our analysis methods, we compare the results using three different forms of $f_{\rm{IGM}}(z)$ and two different prior distributions for $\Omega_{\rm{b,0}}$. Then we find that the uncertainty in $H_0$ is not significantly affected by the prior range of $f_{\rm{IGM}}(z)$ and $\Omega_{\rm{b,0}}$, but the mean value is influenced by the priors for $f_{\rm{IGM}}(z)$ and $\Omega_{\rm{b,0}}$ due to parameter degeneracies with $H_0$. Employing $f_{\rm{IGM}}(z)$ that evolves with redshift, we obtain the constraints for $H_0=69.0^{+6.7}_{-5.7}$ ${\rm{km~s^{-1}~Mpc^{-1}}}$. Furthermore, the mock analyses give a posterior estimation of $H_0$ with an accuracy of 4.6\% and higher precision for $q_0$ and $j_0$ in the near future.
\end{abstract}


\section{Introduction}\label{sec:intro}
The Hubble parameter $H(z)$, characterizes the expansion rate of our universe through its definition using the scale factor and its derivative with respect to cosmic time. At local scales where $z=0$, the Hubble constant $H_0$, plays a critical role as it represents the regression speed of the universe at the present epoch. Therefore, $H_0$ is a crucial cosmological parameter that offers essential insights into our understanding of our universe's behavior.

Nevertheless, with the improved precision in the observation of the distance ladder of the late universe (\cite{Riess:2020fzl,Riess:2021jrx}), cosmic microwave background radiation (CMB), and baryon acoustic oscillation (BAO) (\cite{Abbott:2017wau,Alam:2016hwk,Hildebrandt:2018yau,Hikage:2018qbn}), the application of diverse datasets to constrain $H_0$ results in varying outcomes. The Cepheid-calibrated SNe Ia from SH0ES (\cite{Riess:2021jrx}) found ${H}_0 = 73.04\pm 1.04$ ${\rm{km~s^{-1}~Mpc^{-1}}}$, whilst ${H}_0 = 69.8\pm1.9 $ ${\rm{km~s^{-1}~Mpc^{-1}}}$ when the local distance ladder is calibrated by the tip of the red giant branch (TRGB) methods (\cite{Freedman:2021ahq}), are significantly higher than ${H}_0 = 67.4\pm0.5 $ ${\rm{km~s^{-1}~Mpc^{-1}}}$ and ${H}_0 = 67.6\pm0.5 $ ${\rm{km~s^{-1}~Mpc^{-1}}}$ that inferred from the CMB (\cite{Planck:2018vyg, Chen:2018dbv}) and BAO (\cite{Alam:2016hwk,Gil-Marin:2018cgo,Bautista:2017zgn}), and both of the latter two relies on the determination of the sound horizon scale at either the last scattering surface ($r_s(z_*)$) or the drag epoch ($r_s(z_d)$), leading to the well-known ${H}_0$ tension (\cite{DiValentino:2021izs, Perivolaropoulos:2021jda, Abdalla:2022yfr}).

Concerning the $H_{0}$ tension, the possible solutions can usually be divided into three categories: early-time models \citep{2019PhRvL.122v1301P,2021PhRvD.103f3502M,2020PhRvD.102d3507H}, late-time models \citep{2022MNRAS.511..595Z,2019ApJ...883L...3L,2020MNRAS.497.1590H} and modified gravity \citep{2018JCAP...05..052N,2020PhRvD.101j3505D,2021PhRvD.103b3530A,2022PDU....3500980A}. Recnetly, \citet{2023Univ....9...94H} proposed that according to the path to resolve the $H_0$ tension, the corresponding solutions can be simply divided into two categories, i.e. sequential schemes \citep{2019EPJC...79..576K,2020PhRvD.102b3529B,2020JCAP...01..045X,2021PhRvD.103l1302C,2022PhRvD.105f3535K} and reverse-order schemes \citep{2021ApJ...912..150D,2021PhRvD.104l3511P,2022MNRAS.517..576H,2022PhRvD.106d1301O,2023A&A...674A..45J}. More detail descriptions of the $H_{0}$ tension solutions and its classifications can be found from reviews \citet{2021CQGra..38o3001D}, \citet{2022NewAR..9501659P} and \citet{2023Univ....9...94H}. All in all, although many extensions of $\Lambda$CDM model can alleviate the $H_{0}$ tension, none are supported by the majority of observations. So, hope remains to arbitrate the $H_{0}$ tension using other measurements, for example FRBs.

Fast Radio Bursts (FRBs) are transient radio pulses with burst times in milliseconds and extremely high brightness. Magnetars have emerged as a potential source of FRBs through observational studies (\cite{Bochenek:2020zxn}), but it is worth noting that the production of FRBs may involve other mechanisms as well. Although the exact nature of FRB progenitors remains inconclusive, the dispersion measurement (DM) and redshift relationship make it possible to use extragalactic FRBs for cosmological research (\cite{Macquart:2020lln,Wu:2020jmx,James:2022dcx,Kumar:2019qhc,Zhao:2020ole,Liu:2022bmn,Gao:2022ifq,Zhao:2022yiv}).

The first FRB was discovered by the Parkes Telescope in Australia in 2007 (\cite{Lorimer:2007qn,Chatterjee:2017dqg}). Later, the first repeating FRB121102 discovered by Very Large Array (VLA) (\cite{Spitler:2016dmz,Scholz:2016rpt}) was localized to a dwarf galaxy at $z = 0.19$ after some years (\cite{Chatterjee:2017dqg,Tendulkar:2017vuq,Marcote:2017wan}). In the past few years, the observation of FRBs has undergone significant advancements, thanks to the involvement of projects like Canadian Hydrogen Intensity Mapping Experiment (CHIME) (\cite{CHIMEFRB:2019ghr,CHIMEFRB:2019pgo,Fonseca:2020cdd}), Australian Square Kilometre Array Pathfinder (ASKAP) (\cite{Bannister:2019iju,Bhandari:2020oyb}) and Five hundred-meter Aperture Spherical Telescope (FAST) (\cite{Nan:2011um}). A plethora of FRBs have been detected and their corresponding host galaxies have been identified, leading to a deeper comprehension of the statistical properties and radiation mechanisms associated with FRBs.

The cosmographic approach is a cosmic kinematics approach that defines the Hubble constant ${H}_0$, deceleration parameter $q_0$, jerk parameter $j_0$ and other higher order parameters through the each-order derivatives of the scale factors (\cite{Chiba:1998tc,Visser:2003vq,Visser:2004bf,Capozziello:2013wha,Dunsby:2015ers}). It only takes the cosmological principle FLRW metric as the minimum input and does not depend on the cosmological model. By using the relation between cosmic time, scale factor and redshift, we can expand the observation-related Hubble parameter ${H}(z)$, luminosity  distance $d_{\rm{L}}(z)$ and angular diameter distance $d_{\rm{A}}(z)$ to constrain cosmographic parameters (\cite{Cattoen:2008th}). And these cosmographic parameters are related to the parameters in the cosmological model. The cosmographic approach has been widely used to constrain the curvature of space (\cite{Li:2019qic,Liu:2022lqw}), study the properties of dark energy (\cite{Luongo:2013rba,Luongo:2015zgq}), modified gravity (\cite{Aviles:2012ir,Aviles:2013nga}) and other cosmologic parameters (\cite{Xu:2010hq,Aviles:2012ay,Aviles:2016wel}).

In this work, we employ the cosmographic approach to expand the $\langle{\rm{DM}}_{\rm{IGM}}(z)\rangle$ of FRBs using a Taylor series. This yields a parameterized form that is derived solely from the cosmological principle and is independent of any cosmological model. We use a third-order expansion for parameter constraints. To strengthen the impact of FRBs on constraining the Hubble constant ${H}_0$, we employ a joint constraint approach to cosmographic parameters $q_0$ and $j_0$, utilizing uncalibrated SNe Ia datasets. In this way, we can obtain an posterior distribution of ${H}_0$ through FRB constraints.

The outline of this paper is as follows: in Sec. \ref{sec:cosmography}, we constructed the Hubble parameter ${H}(z)$, luminosity distance $d_{\rm{L}}(z)$ and the mean of dispersion measure of the intergalactic medium $\langle{\rm{DM}}_{\rm{IGM}}(z)\rangle$ based on cosmographic approach, respectively; in Sec. \ref{sec:likelihood}, we present the theory and observation of SNe Ia and FRBs; in Sec. \ref{sec:result}, we present the numerical results obtained using the Markov Chain Monte Carlo (MCMC) method and analyze the results; in Sec. \ref{sec:conclusions}, we summarize our work.
\section{Cosmographic Approach}\label{sec:cosmography}
\begin{figure*}
	\centering
	\includegraphics[scale=0.5]{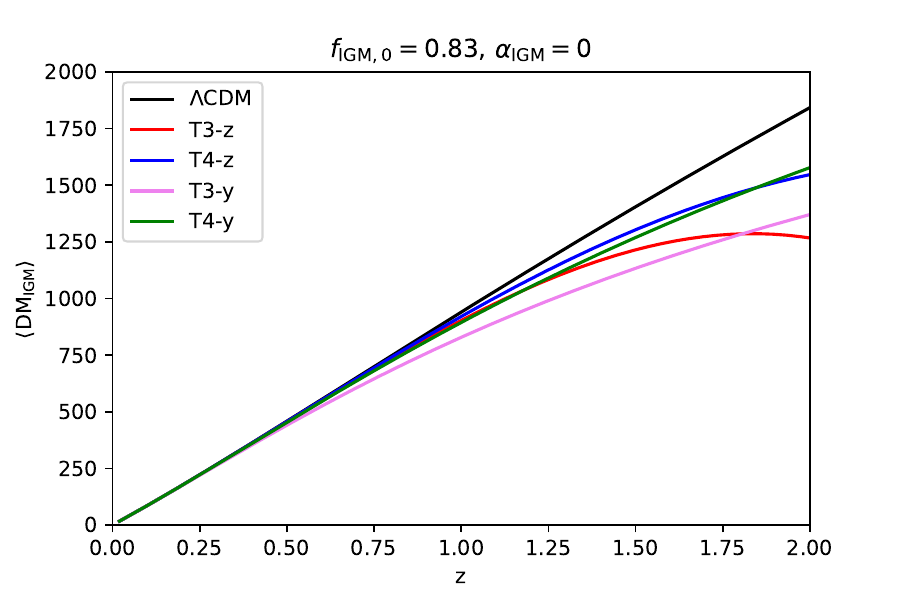}
	\includegraphics[scale=0.5]{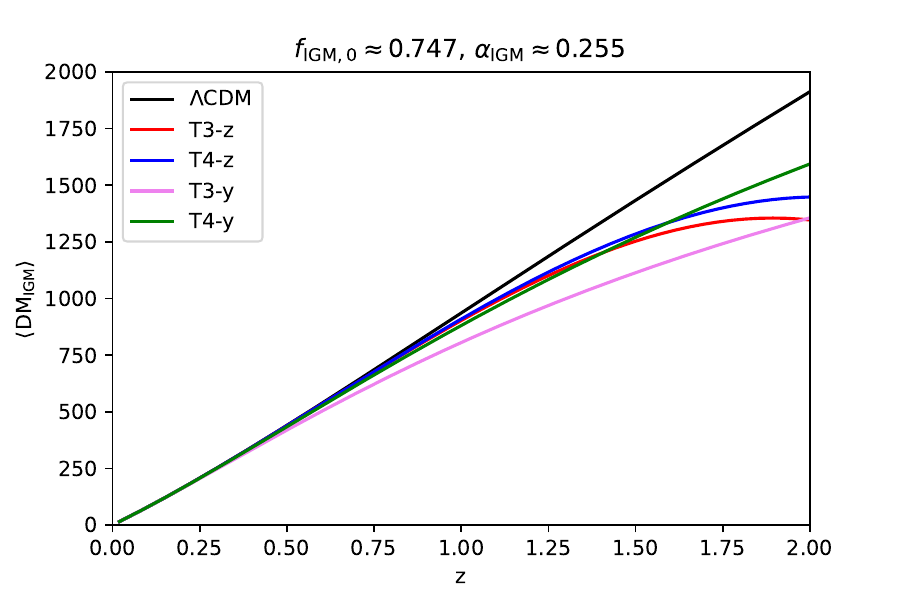}
	\caption{The figure displays the outcome of the Taylor expansion of $\langle{\rm{DM}}_{\rm{IGM}}\rangle$ by adopting the cosmographic approach, utilizing polynomials with $z$ and $y=z/(1+z)$ as the expansion bases. The third-order and fourth-order expansion formulas are represented by T3 and T4, correspondingly. Due to the inherent uncertainty associated with the $f_{\rm{IGM}}(z)$, we have considered it necessary to examine two distinct scenarios. The left plot depicts the implications of case (i) as presented in Sec. \ref{sec:bayesian}, while the right plot illustrates the consequences of case (ii), also outlined in Sec. \ref{sec:bayesian}. Through comparison with the $\Lambda {\rm{CDM}}$ model, the figure depicts the approximation proficiency of various expansion bases and orders of expansion formulas in reference to the $\Lambda {\rm{CDM}}$ model.}
	\label{fig:cosmography}
\end{figure*}
In this section, we present the forms of the Hubble parameter ${H}(z)$, luminosity distance $d_{\rm{L}}(z)$ and mean value of the dispersion measure of the intergalactic medium $\langle{\rm{DM}}_{\rm{IGM}}(z)\rangle$ obtained by using the cosmographic approach.

The cosmographic approach only considers the kinematic equations of cosmological evolution without relying on dynamics. Therefore, there is no need to assume any compositional information about matter, radiation, and dark energy. The approach solely requires the fundamental geometric assumptions of cosmology, and the FLRW metric can be written as
\begin{equation}
	\mathrm{d}s^2=-c^2\mathrm{d}t^2+a^2(t)\left[\frac{\mathrm{d}r^2}{1-kr^2}+r^2(\mathrm{d}\theta^2+\sin^2\theta\mathrm{d}\phi^2)\right],
	\label{eq:FLRW}
\end{equation}
where $c$ represent the velocity of light, $k$ is the curvature of space, and the values 1, 0, $-1$ correspond to the closed, flat and open spatial geometries of the universe, respectively. Recent observations of the CMB, gravitational lensing and BAO, indicate that the spatial curvature can be constrained to $\Omega_{\rm{K}} = 0.0007^{+0.0037}_{-0.0037}$ (\cite{Planck:2018vyg}), or suggest that our universe is flat on large scales. In this paper, only the flat background space with $k=0$ is considered, and different derivatives of the scale factor can be defined as
\begin{eqnarray}
		&H_0\equiv\displaystyle{\frac{\dot{a}}{a}}\bigg|_{0},\quad
		&q_0\equiv-\displaystyle{\frac{1}{{H}^2}}\displaystyle{\frac{\ddot{a}}{a}}\bigg|_{0},\quad \nonumber\\
        &j_0\equiv\displaystyle{\frac{1}{{H}^3}}\displaystyle{\frac{a^{(3)}}{a}}\bigg|_{0},\quad
		&s_0\equiv\displaystyle{\frac{1}{{H}^4}}\displaystyle{\frac{a^{(4)}}{a}}\bigg|_{0},
\end{eqnarray}
where the subscript '0' denotes the value at the present epoch, the dots represent derivative with respect to cosmic time and $a^{(n)}$ stands for the $n$-th time derivative of scale factor $a$. Using the relationship between cosmic time and redshift $\frac{\mathrm{d} t}{\mathrm{d}z}=-\frac{1}{(1+z)H(z)}$, the derivatives of the Hubble parameter with respect to redshift $z$ can be calculated as
\begin{subequations}
	\begin{align}
		\frac{\mathrm{d}H}{\mathrm{d}z}\bigg|_0&=(1+q_0)H_0,\label{eq:h1} \\
		\frac{\mathrm{d}^2H}{\mathrm{d}z^2}\bigg|_0&=(j_0-q_0^2)H_0,\label{eq:h2}	\\
		\frac{\mathrm{d}^3H}{\mathrm{d}z^3}\bigg|_0&=[3q_0^3+3q_0^2-j_0(3+4q_0)-s_0]H_0.\label{eq:h3}
	\end{align}
\end{subequations}
Based on the order expressions of the Hubble parameter (Eq.~(\ref{eq:h1}), Eq.~(\ref{eq:h2}) and Eq.~(\ref{eq:h3})), the Taylor expansion of the Hubble parameter ${H}(z)$ can be directly written as
\begin{equation}
	\begin{aligned}
	H(z) &=H_0\{1+(1+q_0)z+(j_0-q_0^{2})z^2/2+[3q_{0}^{3}+3q_{0}^{2}\\&-j_0(3+4q_0)-s_0]z^3/6+\cdots\}.
	\label{eq:hz}
	\end{aligned}
\end{equation} 

According to the comoving distance $d_{\rm{c}}(z)=\int^{z}_{0}\frac{c}{{H}(z^{\prime})}\mathrm{d}{z^{\prime}}$ and the dual relationship between the luminosity distance and comoving distance $d_{\rm{L}}(z)= (1+z)d_{\rm{c}}(z)$, the Taylor expansion of the luminosity distance $d_{\rm{L}}(z)$ can be calculated as
\begin{equation}
	\begin{aligned}
	d_{\rm{L}}(z)&=cH_0^{-1}\{z+(1-q_0)z^2/2-(1-q_0-3q_0^2+j_0)\\&z^3/6+[2-2q_0-15q_0^2-15q_0^3+5j_0+10q_0j_0\\&+s_0]z^4/24+\cdots\}.
	\label{eq:dlz}
	\end{aligned}
\end{equation}
In light of Eq.~(\ref{eq:dmigm}), one can Taylor expand the $\langle{\rm{DM}}_{\rm{IGM}}(z)\rangle$ of FRBs as
\begin{equation}
	\begin{aligned}
		\langle{\rm{DM}}_{\rm{IGM}}(z)\rangle&=\frac{3c\Omega_{\rm{b,0}}{H_0}\chi_e}{8\pi Gm_p}\{f_0z+(\alpha-f_0q_0)z^2/2\\&+[-f_0j_0-2\alpha(1+q_0)+f_0q_0(2+3q_0)]\\&z^3/6+\{\alpha(6-3j_0+12q_0+9q_0^2)+f_0[-6q_0\\&-18q_0^2-15q_0^3+2j_0(3+5q_0)+s_0]\}z^4/24\\&+\cdots\},
		\label{eq:dmz}
	\end{aligned}
\end{equation}
where $f_{0}$ is shorthand for $f_{\rm{IGM,0}}$, and $\alpha$ is shorthand for $\alpha_{\rm{IGM}}$. The Taylor series Eq.~(\ref{eq:hz}), Eq.~(\ref{eq:dlz}), and Eq.~(\ref{eq:dmz}) are not convergent when $z>1$. However, by utilizing the parameter substitution $y=z/(1+z)$, we are able to obtain the convergent Taylor series of ${H}(y)$, $d_{\rm{L}}(y)$, and $\langle{\rm{DM}}_{\rm{IGM}}(y)\rangle$. The resulting Taylor expansions can be expressed as
\begin{equation}
	\begin{split}
		\begin{aligned}
			H(y)&=H_0[1+(1+q_0)y+(1+q_0+j_0/2-q_0^2/2)y^2\\&+(6+3j_0+6q_0-4q_0j_0-3q_0^2+3q_0^3-s_0)y^3/6\\
			&+(1+q_0-2j_0q_0+3q_0^3/2-s_0/2)y^4+\cdots],
		\end{aligned}
	\end{split}
	\label{eq:hy}
\end{equation}
\begin{equation}
	\begin{split}
		\begin{aligned}
			d_{\rm{L}}(y)&=cH_0^{-1}[y+(3-q_0)y^2/2+(11-j_0-5q_0\\&+3q_0^2)y^3/6+(50-7j_0-26q_0+10q_0j_0+21q_0^2 \\
			&-15q_0^3+s_0)y^4/24+\cdots],
		\end{aligned}
	\end{split}
	\label{eq:dly}
\end{equation}
\begin{equation}
	\begin{split}
		\begin{aligned}
			\langle{\rm{DM}}_{\rm{IGM}}(y)\rangle&=\frac{3c\Omega_{\rm{b,0}}{H_0}\chi_e}{8\pi Gm_p}\{f_{\rm{0}}y+(\alpha+2f_{\rm{0}}-f_{\rm{0}}q_0)y^2/2\\&-[2\alpha(-2+q_0)+f_{\rm{0}}(-6+j_0+4q-3q^2)]\\&y^3/6+\{-3\alpha(-6+j_0+4q_0-3q_0^2)+f_{\rm{0}}[24\\&-18q_0+18q_0^2-15q_0^3+2j_0(-3+5q_0)\\&+s_0]\}y^4/24+\cdots\}.
		\end{aligned}
	\end{split}
	\label{eq:dmy}
\end{equation}

To illustrate the capabilities of the comgraphic approach, we can compare the third-order and fourth-order of the Taylor expansion Eq.~(\ref{eq:dmz}) and Eq. (\ref{eq:dmy}) with the $\Lambda {\rm{CDM}}$ model. At the late time of our universe, the $\Lambda {\rm{CDM}}$ model can be represented as
\begin{equation}
	{H}(z)={H}_0\sqrt{\Omega_{\rm{m}}(1+z)^3+\Omega_{\rm{DE}}}, 
\end{equation}
where $\Omega_{\rm{m}}$ is the dimensionless density of matter, and $\Omega_{\rm{DE}}$ is the dimensionless density of dark energy. The corresponding relationship between cosmographic parameters and flat $\Lambda$CDM model parameters is $q_0=\frac{3}{2}\Omega_{\rm{m}}-1$, $j_0=1$, $s_0=1-\frac{9}{2}\Omega_{\rm{m}}$ (\cite{Li:2019qic,Xu:2010hq}).

In this work, we adopt the values of ${H}_0=70$ ${\rm{km~s^{-1}~Mpc^{-1}}}$ and $\Omega_{\rm{m}}=0.3$, along with the corresponding cosmographic parameters of ${H}_0=70$ ${\rm{km~s^{-1}~Mpc^{-1}}}$, ${q}_0=-0.55$, ${j}_0=1$, and ${s}_0=-0.35$, as presented in Fig. \ref{fig:cosmography}. Due to the uncertainty of $f_{\rm{IGM}}(z)$, here we compare $\Lambda$CDM model with both case (i) and case (ii) from Sec. \ref{sec:bayesian} to illustrate the approximation ability of the cosmographic approach. The left panel in Fig. \ref{fig:cosmography} shows case (i), where we adopt $f_{\rm{IGM,0}}(f_0)=0.83$ and $\alpha_{\rm{IGM}}(\alpha)=0$. The right panel in Fig. \ref{fig:cosmography} shows case (ii), where $f_{\rm{IGM}}(z)$ evolves with redshift and we adopt $f_{\rm{IGM,0}}(f_0)\approx0.747$ and $\alpha_{\rm{IGM}}(\alpha)\approx0.255$. In the two cases, the cosmographic approach shows a similar trend to that of the $\Lambda$CDM model. Our results demonstrate that various expansions utilized in the redshift range of the current FRBs data ($z<0.66$) exhibit small deviations from the $\Lambda$CDM model. Specifically, the Taylor expansion in terms of $z$ is especially effective when $z<1$.

In the case of future FRBs located at higher redshifts (\cite{Hashimoto:2020dud}), it is essential to exercise caution when applying cosmographic approach. When $z>1$, the Taylor expansion using $z$ (Eq. (\ref{eq:dmz})), which effectively approximates distances at lower redshifts, may not converge, thereby rendering it unreliable. In contrast, the Taylor expansion using $y$ (Eq. (\ref{eq:dmy})) guarantees convergence even at higher redshifts. However, Fig. \ref{fig:cosmography} demonstrates that when $z>0.5$, the third-order of Eq. (\ref{eq:dmy}) begins to deviate from the $\Lambda$CDM model, leading to discrepancies between the cosmographic parameters and the corresponding physical parameters in the $\Lambda$CDM model.

One potential solution to suppress the deviation is to increase the order of the Taylor expansion. However, this approach results in a larger parameter space, introducing new challenges. Compared to the Hubble parameter, luminosity distance and angular diameter distance (\citet{Li:2019qic}), the Taylor expansion of $\langle {\rm{DM}}_{\rm{IGM}} \rangle$ obtained through the cosmographic approach demonstrates significantly larger deviations from the $\Lambda$CDM model at earlier epochs. It is hypothesized that the discrepancy may be attributed to the distinctive form of $\langle {\rm{DM}}_{\rm{IGM}} \rangle$. Alternative forms, such as Padé polynomials (\cite{Cattoen:2007sk,Aviles:2014rma,Gruber:2013wua}), may potentially address this issue at higher redshifts, and will be discussed in further analysis.

\section{Observations and Theories}\label{sec:likelihood}
Even when using the third-order form described as Eq.~(\ref{eq:dmigm}) to approximate $\langle{\rm{DM}}_{\rm{IGM}}(z)\rangle$, there are still three cosmographic parameters: ${H}_{0}$, $q_0$ and $j_0$. With the current observational FRBs data, it is difficult to constrain these three cosmographic parameters . To address this issue, we consider uncalibrated SNe Ia data, which can help constrain the high-order parameters $q_0$ and $j_0$ and avoid the use of the ${H}_{0}$ prior.
\subsection{SNe Ia}
SNe Ia is a kind of standard candle, the observed brightness of SNe Ia depends on the luminosity distance $d_{\rm{L}}$, and the quantity describing the apparent brightness is the apparent magnitude, which can be written as
\begin{equation}
	{m}=5{\rm{log}}({d_{\rm{L}}}(z))+{M_B}+25,
\end{equation}
where the ${d_{\rm{L}}}$ is the luminosity distance $[{\rm{Mpc}}]$, the ${M_B}$ is the absolute magnitude.

In the case of an unknown absolute magnitude, we can also define the distance modulus for cosmological parameter constraints, the distance modulus is defined as
\begin{equation}
	{\mu}_{\rm{th}}(z)\equiv {m}_{\rm{th}}(z,\{\theta\})-{M}_{B}=5{\rm{log}}({\widetilde{d_{\rm{L}}}}(z,\{\theta\}))+\mu_0,
\end{equation}
where the $\widetilde{d_{\rm{L}}}(z)={H}_{0}d_{\rm{L}}(z)$ is independent of the Hubble constant ${H}_{0}$, while the ${H}_{0}$ dependency is encapsulated in parameter $\mu_0= 42.38-5{\rm{{log}_{10}}}h$, where the $h\equiv{H}_0/100$. Consequently, in the Pantheon samples where $\mu_0$ is marginalized over, $H_0$ is unconstrained. 

In this paper, we use the Pantheon sample, which contains 1048 SNe spanning the redshift range $0.01 < z < 2.3$ (\cite{Pan-STARRS1:2017jku}) and not dependent on any cosmological model (\cite{Kessler:2016uwi}) and is used to constrain only $q_0$ and $j_0$. 

The likelihood function of Pantheon data is written as
\begin{equation}
	\chi_{\rm{Pantheon}}^2(\{\theta\},M^{\prime})=\Delta\hat{\mu}^{\rm{T}}\cdot Cov^{-1}\cdot\Delta\hat{\mu},
	\label{eq:lsn}	
\end{equation}
where $\Delta\hat{\mu}_{i}=m_{B,i}^{*}-5\mathrm{log}_{10}[d_L(z_i)]+(M^{\prime}-\mu_0)$, $m_{B,i}^{*}$ and $M^{\prime}$ are the observations, and $M^{\prime}-\mu_0$ is marginalized over analytucally (\cite{DiPietro:2002cz,Nesseris:2004wj,Nesseris:2005ur}). In previous work, SNe Ia data was often combined with Cepheids (\cite{Riess:2021jrx}) or TRGB (\cite{Freedman:2021ahq}) dataset to constrain ${H}_{0}$. However, in this work, we use FRBs dataset as a calibrator for SNe Ia to constrain ${H}_{0}$ and the late expansion history.

\subsection{FRBs}
Fast Radio Bursts (FRBs) are used to constrain cosmological parameters,  depend on the relationship between Dispersion Measure (DM) and redshift. The high-intensity electromagnetic pulses generated by FRBs are dispersed by plasma dispersion along their propagation path, causing signals of different frequencies to arrive at different times. DM can be obtained by processing FRB signals in the frequency domain (\cite{Deng:2013aga,Petroff:2019tty}), which can be written as
\begin{equation}
	\Delta t\propto\left(\frac{1}{v_{1}^{2}}-\frac{1}{v_{2}^{2}}\right){\rm{DM}}.
	\label{eq:dmobs}
\end{equation}

The full-path integral of plasma, denoted as ${\rm{DM}}=\int n_e/(1+z)~\mathrm{d}l$, is used to consider the cosmological influences on ${\rm{DM}}$. This represents the total ${\rm{DM}}$ of extragalactic FRBs is written as
\begin{equation}
	{\rm{DM}}_{\rm{FRB}}={\rm{DM}}_{\rm{ISM}}+{\rm{DM}}_{\rm{halo}}+{\rm{DM}}_{\rm{IGM}}+\frac{{\rm{DM}}_{\rm{host}}}{(1+z)}.
	\label{eq:dm}
\end{equation}

In Eq.~(\ref{eq:dm}), we denote the contributions from plasma accumulated in the interstellar medium (ISM) and halo of the Milky Way by ${\rm{DM}}_{\rm{ISM}}$ and ${\rm{DM}}_{\rm{halo}}$, respectively. ${\rm{DM}}_{\rm{ISM}}$ is mainly produced by the contribution of warm ionized medium, and it can be calculated with Milky Way disk-shaped free electron distribution models, such as ${\rm{NE}}2001$ (\cite{Cordes:2002wz}) and ${\rm{YMW}}16$ (\cite{yao2017new}), which are based on astronomical observations. In this paper, we use ${\rm{NE}}2001$ to compute the value of ${\rm{DM}}_{\rm{ISM}}$. ${\rm{DM}}_{\rm{halo}}$ is used to represent the contributions originating from the extended hot Galactic halo, which includes a spherical component of isothermal gas and disk-like non-spherical hot gas in the Milky Way and its edges. Refs. \citet{Dolag:2014bca}, \citet{Yamasaki:2019htx}, and \citet{prochaska2019probing} have previously discussed this part. In this papar, we follow \citet{Macquart:2020lln} and \citet{Lin:2023opv} set ${\rm{DM}}_{\rm{halo}} = 50$ ${\rm{pc}}\cdot {\rm{cm}}^{-3}$

${\rm{DM}}_{\rm{IGM}}$ represents the dispersion measurement in the intergalactic medium (IGM) of the large scale structure. The mean of ${\rm{DM}}_{\rm{IGM}}$ at redshift $z$ can be written as
\begin{equation}
	\langle{\rm{DM}}_{\rm{IGM}}(z)\rangle=\frac{3c\Omega_{\rm{b,0}} {H}_{\rm{0}}}{8\pi Gm_p}\int^{z}_{0}\frac{f_{\rm{IGM}}(z^{\prime})\chi_{\rm{e}}(z^{\prime})(1+z^{\prime})}{E(z^{\prime})}dz^{\prime},
	\label{eq:dmigm}
\end{equation}
where $\Omega_{\rm{b,0}}$, $G$ and $m_{\rm{p}}$ represent the baryon density, gravitational constant, and the mass of the proton, respectively. We have the dimensionless baryon density $\Omega_{\rm{b,0}} = 0.0487^{+0.0005}_{-0.0004}$ to the best-fit values found in Ref. \citet{DES:2021wwk}. $\chi_{\rm{e}}=Y_{\rm{H}}\chi_{\rm{e,H}}+1/2Y_{\rm{He}}\chi_{\rm{e,He}}=7/8$ is electron fraction, where $Y_{\rm{H}}=3/4$ and $Y_{\rm{He}}=1/4$ are the mass fractions of hydrogen and helium, respectively. We assume that $\chi_{\rm{e,H}}$ and $\chi_{\rm{e,He}}$, the ionization fractions of intergalactic hydrogen and helium, are fully ionized at $z<3$ (\cite{Meiksin:2007rz,becker2011detection}). $f_{\rm{IGM}}(z)$ is the fraction of baryons in the IGM, and the uncertainty remains large around its value. When performing a series expansion of $\langle{\rm{DM}}_{\rm{IGM}}(z)\rangle$ using the cosmographic approach, we consider a general form $f_{\rm{IGM}}(z)=f_{\rm{IGM,0}}+\alpha_{\rm{IGM}} z/(1+z)$, which evolves linearly with redshift. In the parameter space analysis presented in Sec. \ref{sec:bayesian}, we provide assessments on different scenarios of $f_{\rm{IGM,0}}$ and $\alpha_{\rm{IGM}}$.

\begin{figure}
	\includegraphics[scale=0.5]{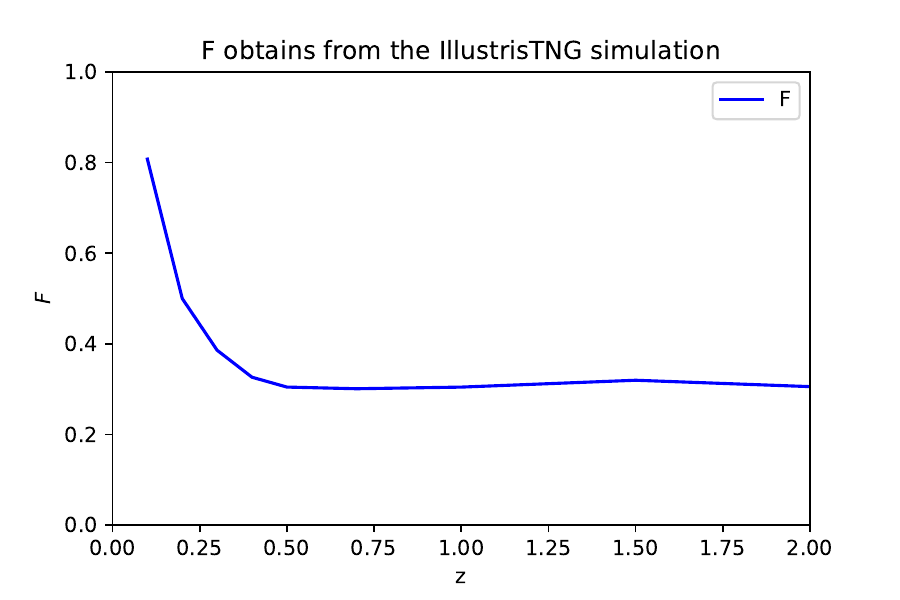}
	\caption{The relationship between $F$ and redshift $z$ derived from the result of the state-of-the-art IllustrisTNG simulation.}
	\label{fig:Fz}
\end{figure}

Due to the existence of galaxies and halos, the distribution of ionized plasma in the intergalactic medium is inhomogeneous. As a result, the actual ${\rm{DM}}_{\rm{IGM}}(z)$ will deviate from $\langle{\rm{DM}}_{\rm{IGM}}(z)\rangle$. We use a probability distribution function of the following theoretical form (\cite{Macquart:2020lln,zhang2021intergalactic}) as
\begin{equation}
	{P}_{\rm{IGM}}(\Delta)=A\Delta^{-\beta}{\rm{exp}}\left[-\frac{(\Delta^{-\alpha}-C_{0})^2}{2\alpha^2\sigma_{\rm{IGM}}^2}\right], \Delta>0,
	\label{eq:pcos}	
\end{equation}
where  $A$ is the normalization factor, $\Delta={\rm{DM}}_{\rm{IGM}}(z_i)/ \langle{\rm{DM}}_{\rm{IGM}}(z)\rangle$, $\alpha=\beta=3$ are related to the inner density profile of gas in halos (\cite{Macquart:2020lln}), $\sigma_{\rm{IGM}}$ represents the effective standard deviation. The mean of the distribution requires that $\langle \Delta\rangle=1$, which fixes the free parameter $C_0$ in $P_{\rm{IGM}}(\Delta)$. Ref. \citet{Macquart:2020lln} proposed a form $\sigma_{\rm{IGM}}=Fz^{-0.5}$ that uses a constant $F$ to describe the error and participates in MCMC sampling as a free parameter, but there is a possibility that $F$ varies with redshift and changes significantly in certain redshift ranges. In this case, the posterior distribution of $F$ obtained using the Macquart's relation may strongly depend on the redshift distribution of the data and be degenerate with other parameters that are then passed on to cosmological parameters. Fig. \ref{fig:Fz} illustrates the relationship between $F$ and redshift $z$ derived from the result of the state-of-the-art IllustrisTNG simulation in Ref. \citet{Zhang:2020xoc}. It reveals that $F$ may be not a constant at low redshifts. In this paper, we follow Ref. \citet{wu20228} and use the results of the state-of-the-art IllustrisTNG simulation in Ref. \citet{Zhang:2020xoc} to estimate $\sigma_{\rm{IGM}}$ and $C_0$ at different redshift for each FRB.

${\rm{DM}}_{\rm{host}}$ represents the contribution from the host galaxy and the source environment of FRBs. IllustrisTNG simulations of three possible origins provide a long-tailed distribution prior for the host galaxy (\cite{Zhang:2020mgq}) is represented by
\begin{equation}
	\begin{small}
	\begin{aligned}
	{P}_{\rm{host}}({\rm{DM}}_{\rm{host}}\textbar \mu,\sigma_{\rm{host}})=\frac{1}{\sqrt{2\pi}{\rm{DM}}_{\rm{host}}\sigma_{\rm{host}}}{\rm{exp}}\left[-\frac{({\rm{ln}}{\rm{DM}}_{\rm{host}}-\mu)^2}{2\sigma_{\rm{host}}^2}\right],
	\end{aligned}
	\end{small}
\end{equation}
where $\mu$ and $\sigma_{\rm{host}}$ are the median and standard deviation of ${\rm{ln}}{\rm{DM}}_{\rm{host}}$, respectively. Based on three possible types of FRBs included repeating FRBs like FRB 121102, repeating FRBs like FRB 180916 and non-repeating FRBs individually, ${\rm{DM}}_{\rm{host}}$ is evolves with redshift (\cite{Zhang:2020mgq}) and the median of ${\rm{DM}}^{\prime}_{\rm{host}}(z)=A_{\rm{host}}(1+z)^{\alpha_{\rm{host}}}$, where the ${\rm{DM}}^{\prime}_{\rm{host}}={\rm{DM}}_{\rm{host}}/(1+z)$. Here we classify the localized FRBs based on Ref. \citet{Zhang:2020mgq} and provide the parameters $\{A_{i}, \alpha_{i}\}$, where the subscripts $"i"$ here can take values 1, 2, 3, corresponding to the values of the three types of FRBs. Based on the results of the IllustrisTNG simulation, $\sigma_{\rm{host}}$ also depends on the type and redshift of FRBs. Here, we use the results of Ref. \citet{Zhang:2020mgq} and interpolate to calculate the $\sigma_{\rm{host}}$ of each FRB.

Therefore, the distribution function of the Extragalactic dispersion measurement can be determined using ${P}_{\rm{IGM}}(\Delta)$ and ${P}_{\rm{host}}({\rm{DM}}_{\rm{host}}\textbar \mu,\sigma_{\rm{host}})$ as
\begin{equation}
	\begin{small}
	\begin{aligned}
		{P}_{\rm{EG}}({\rm{DM}}_{\rm{EG}})
		&=\int_{0}^{{\rm{DM}}_{\rm{EG}}} {P}_{\rm{IGM}}({\rm{DM}}_{\rm{EG}}-{\rm{DM}^{\prime}_{\rm{host}}})
		\\&\times{P}_{\rm{host}^{\prime}}({\rm{DM}}^{\prime}_{\rm{host}}){\rm{d}}{\rm{DM}_{\rm{host}}^{\prime}},
	\end{aligned}
    \end{small}
\end{equation}
where ${\rm{DM}}_{\rm{EG}}={\rm{DM}}_{\rm{IGM}}+{\rm{DM}}^{\prime}_{\rm{host}}={\rm{DM}}_{\rm{obs}}-{\rm{DM}}_{\rm{ISM}}-{\rm{DM}}_{\rm{halo}}$, ${\rm{DM}}_{\rm{obs}}$ is the observation of FRBs, and ${\rm{DM}}_{\rm{ISM}}$ is calculated by NE2001 model. For the localized FRBs, with ${\rm{DM}}_{\rm{IGM}}$ and ${\rm{DM}}^{\prime}_{\rm{host}}$ as independent variables, when calculating ${P}_{\rm{EG}}({\rm{DM}}_{\rm{EG}})$, we normalize ${P}_{\rm{IGM}}(\rm{DM}_{\rm{IGM}})$ and ${P}_{\rm{host}^{\prime}}({\rm{DM}}^{\prime}_{\rm{host}})$ with respect to $\rm{DM}_{\rm{IGM}}$ and ${\rm{DM}}^{\prime}_{\rm{host}}$ respectively. Since $\rm{DM}_{\rm{IGM}}>0$, we set the upper limit of integration for ${\rm{DM}}^{\prime}_{\rm{host}}$ as ${\rm{DM}}_{\rm{EG}}$.

Regarding all the FRBs data, the joint likelihood function is written as
\begin{equation}
	\mathcal{L}=\prod_{i=1}^{N}{P}_{\rm{FRB}}({\rm{DM}}_{\rm{FRB},i}\textbar z_{i}),
	\label{eq:lfrb}
\end{equation} 
where the subscript $"i"$ here represents the $i$-th FRBs data. $N$ is the total number of FRBs data. The FRBs data used in this study were obtained from Ref. \citet{wu20228}, which includes 18 localized FRBs (\cite{Chatterjee:2017dqg,Bhandari:2021pvj,Marcote:2020ljw,Bannister:2019iju,Bhardwaj:2021hgc,doi:10.1126/science.aay0073,Bhandari:2020oyb,Ravi:2019alc,Chittidi_2021,Heintz_2020,Law:2020cnm,Ravi:2021kqk}). These 18 localized FRBs can be classified according to their galaxy properties and phenomenological characteristics, and in this paper, we adopt the classification results from Ref. \citet{Yang_2022}.

\subsection{Bayesian analyses}\label{sec:bayesian}

According to Bayesian theory, the parameter probability distribution of the posterior is proportional to the probability distribution of the data, which can be written as
\begin{equation}
	{P}(\{\theta\}\textbar {\rm{data}}) = \frac{\mathcal{L}({\rm{data}}\textbar\{\theta\}){P}(\{\theta\})}{{P}({\rm{data}})},
\end{equation}
where the ${P}(\rm{data})$ is global likelihood function, $\mathcal{L}({\rm{data}}\textbar\{\theta\})$ is the likelihood function, ${P}(\{\theta\})$ is the prior probability distribution of the parameters.

In this paper, we use FRBs data in the redshift range of $(0.0039, 0.66)$ and apply the third-order Taylor expansion Eq. (\ref{eq:dmz}) for posterior estimation. Based on the previous research \citep{2017EPJC...77..434Z,2022A&A...661A..71H}, for the Pantheon data in the redshift range of $(0.1, 2.3)$, we adopt the safest third-order expansion with three free parameters ($H_{0}$, $q_{0}$ and $j_{0}$). The approximation ability of Eq. (\ref{eq:dly}) is demonstrated in Ref. \citet{Li:2019qic}.

We obtain the posterior distribution of parameters using three different combinations of observational data: Pantheon, FRB and Pantheon+FRB. The parameter space $\{\theta\}$ for Pantheon consists of $\{{H}_0, q_0, j_0\}$. For Pantheon+FRB, we will analyze three possible scenarios separately due to the different possibilities and uncertainties of $f_{\rm{IGM}}(z)$. Research on reionization history estimates $f_{\rm{IGM}}(z)$ to be 0.82 and 0.9 at $z<0.4$ and $z>1.5$, respectively (\cite{Shull:2011aa}). Direct observation of baryons provides an approximate value of $f_{\rm{IGM,0}}\approx 0.83$ (\cite{fukugita1998cosmic}). In addition, Ref. \citet{Li:2020qei} used five FRBs constraints to constrain $f_{\rm{IGM}}(z)$, but did not find strong evidence for evolution with redshift. Ref. \citet{Lin:2023opv} used 18 FRBs to constrain $f_{\rm{IGM}}(z)$, but also found it difficult to come to a definitive conclusion for the two free parameters, $f_{\rm{IGM,0}}$ and $\alpha_{\rm{IGM}}$. In this paper, we discuss three possible scenarios: (i) $f_{\rm{IGM}}(z)$ does not evolve with redshift and we assume $f_{\rm{IGM,0}}=0.83$ and $\alpha_{\rm{IGM}}=0$; (ii) $f_{\rm{IGM}}(z)$ evolves with redshift, and we set $f_{\rm{IGM}}(0,4)=0.82$ and $f_{\rm{IGM}}(1.5)=0.9$ to calculate $f_{\rm{IGM,0}}$ and $\alpha_{\rm{IGM}}$; (iii) we treat $f_{\rm{IGM,0}}$ as a free parameter and give a prior interval of $f_{\rm{IGM,0}} \in [0.747,0.913]$. For the first two scenarios, the parameter space is $\{{{H}_0,q_0,j_0,\Omega_{\rm{b,0}}}\}$. For the third scenario, the parameter space is $\{{{H}_0,q_0,j_0,\Omega_{\rm{b,0}},f_{\rm{IGM,0}}}\}$. For the using only FRB data, we set $f_{\rm{IGM}}(z)$ as case (i), with the parameter space is $\{{{H}_0,q_0,j_0,\Omega_{\rm{b,0}}}\}$. Therefore, we can use Eq. (\ref{eq:lsn}) and Eq. (\ref{eq:lfrb}) to construct the likelihood function.

For both types of data, we use the $\texttt{cobaya}$ for Markov Chain Monte Carlo (MCMC) sampling (\cite{torrado2019cobaya,Torrado:2020dgo}), which is an open source code for bayesian analysis in cosmology, and we use the $\texttt{GetDist}$ to analyze the sampled chains (\cite{Lewis:2019xzd}). We will provide the prior distribution of the parameters given for the MCMC sampling and the posterior distribution obtained from the sampling in Sec. {\ref{sec:result}}.

\section{Results and Analysis}\label{sec:result}

We employ the likelihood function and datasets described in Sec. \ref{sec:likelihood} to perform MCMC sampling and obtain the results of the parameters.

To account for potential parameter degeneracies, we sample and compare three possible forms of prior distributions for $f_{\rm{IGM,0}}$. For $\Omega_{\rm{b,0}}$, we only specify the sampling range without a particular prior distribution. The parameter ranges are set as $\Omega_{\rm{b,0}}\in \mathcal{U}(0.0483,0.0492)$ ($1\sigma$) and $\Omega_{\rm{b,0}}\in \mathcal{U}(0.0467,0.0513)$ ($5\sigma$), respectively, to investigate the influence of baryon density fraction uncertainty on $H_0$. In terms of cosmographic parameters, we select larger parameter ranges around the $\Lambda {\rm{CDM}}$ model to explore a wider range of potential parameter combinations, namely ${H}_0\in \mathcal{U}(20,120)$, $q_0\in \mathcal{U}(-2,0)$, $j_0\in \mathcal{U}(-10,10)$. 
	
Furthermore, constraints must be imposed on the physical quantities derived from cosmographic parameters to ensure that they have valid physical interpretations. Therefore, before MCMC sampling, we perform conditional checks on parameter combinations. Specifically, if ${H}(y)$, ${\rm{d}}_{\rm{L}}(y)$, or $\langle{\rm{DM}}_{\rm{IGM}}(z)\rangle$ is negative, the probability is set to zero. To account for the truncation order of the cosmographic approach and the expansion history of the universe, we set the redshift range to $(0,2)$.	
When plotting the probability distribution function of the parameters, it is crucial to consider the convergence of MCMC sampling. In this study, we set $r-1<0.01$ as the convergence criterion to ensure the MCMC sampling has converged.

\subsection{The Constraints of  $H_0$, $q_0$, and $j_0$}\label{sec:result1}

\begin{figure}
	\includegraphics[scale=0.5]{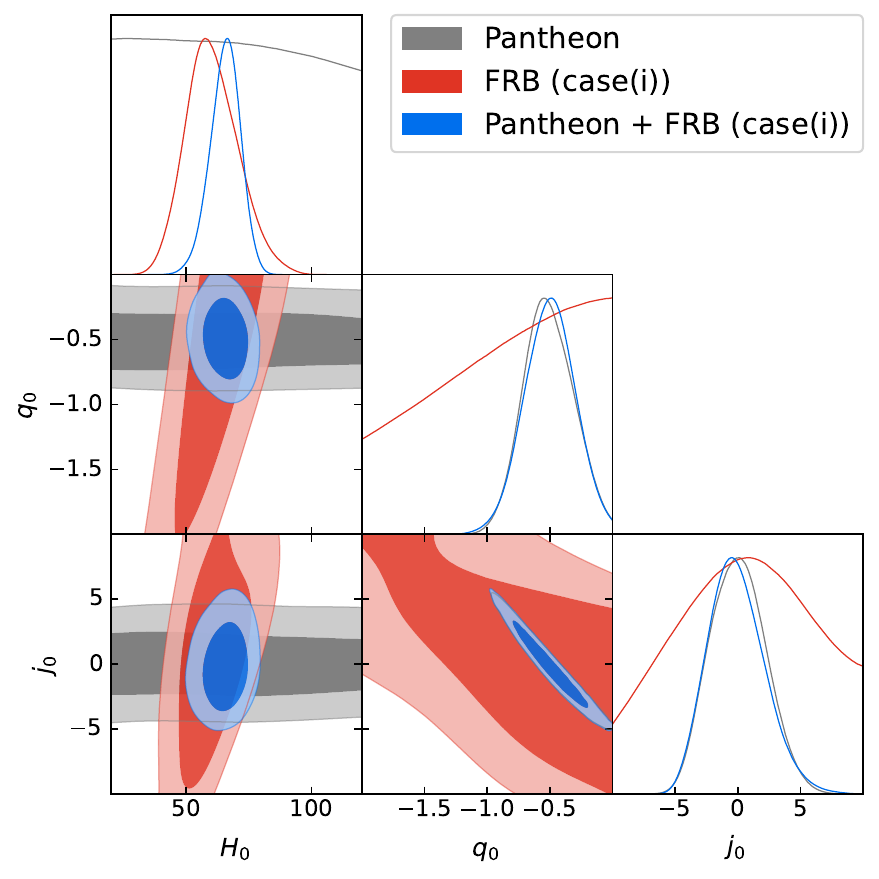}
	\caption{The contour plots include the $68\%$ and $95\%$ confidence intervals constrained by Pantheon, FRBs, and Pantheon+FRB. In the cases of FRBs and Pantheon+FRB, $f_{\rm{IGM}}(z)$ is set as case (i) and $\Omega_{\rm{b,0}}$ is chosen to have a uniform prior with $1 \sigma$ deviation.}
	\label{fig:result1}
\end{figure}

For the measurement of $H_0$, $q_0$, and $j_0$, we sample the posterior distributions of cosmographic parameters using FRB data and the Pantheon dataset separately. In this cases, we use case (i) for $f_{\rm{IGM,0}}$ and $\Omega_{\rm{b,0}}\in \mathcal{U}(0.0483,0.0492)$ ($1\sigma$). Subsequently, we combine the constraints from FRB and Pantheon to obtain the posterior distributions of the cosmographic parameters $\{{H}_0,q_0,j_0\}$, which are presented in Tab. \ref{tab:1}. The contours in Fig. \ref{fig:result1} are plotted using the chains. 

\begin{table}
	\centering
	\caption{The results obtained using Pantheon, FRBs, and Pantheon+FRB constraints separately. In the case of FRB and Pantheon+FRB, $f_{\rm{IGM}}(z)$ is set as case (i) and $\Omega_{\rm{b,0}}$ is chosen to have a uniform prior with $1 \sigma$ deviation.}
	\begin{tabular}{l|c|c|c|c} 
		\hline
		Dataset & ${\rm{H}}_0$& $q_0$ & $j_0$ & $\Omega_{\rm{b,0}}$ \\
		\hline
		Pantheon & --- & $-0.51\pm 0.20$ & $0.0\pm 2.2$& \\
		FRB & $60^{+9}_{-10}             $ & $> -1.13                   $ & $0.5\pm 5.0    $ & ---       \\
		Pantheon+FRB & $65.5^{+6.4}_{-5.4}        $ & $-0.50\pm 0.20             $ & $-0.1^{+2.0}_{-2.5}        $ & ---       \\
		\hline
	\end{tabular}
	\label{tab:1}
\end{table}

The grey contour in Fig. \ref{fig:result1} represents the constraint obtained by using only the Pantheon dataset to determine $\{{H}_0,q_0,j_0\}$. We obtain a constraint result of $q_0=-0.51\pm 0.20$ and $j_0=0.0\pm 2.2$ with $68\%$ confidence level, which provides some insight into the higher-order evolution of the universe. However, due to the marginalization of $M^{\prime}-\mu_0$, the uncalibrated SNe Ia dataset can not effectively constrain $H_0$. This result also supports the validity of the prior constraints on the parameters in this paper.

The red contour in Fig. \ref{fig:result1} represents the results obtained using FRB constraints alone, where $H_0=60^{+9}_{-10}$ ${\rm{km~s^{-1}~Mpc^{-1}}}$. It is evident that it is challenging to constrain the three cosmographic parameters using the 18 FRB data in this paper and the model selection of FRB described in Sec. \ref{sec:likelihood}. From the contour plot, we observe a degeneracy between $H_0$ and $q_0$ parameters, where a smaller $q_0$ allows for a smaller $H_0$ due to this parameter degeneracy. The precision of the FRB constraints on $q_0$ is relatively low in this study, and we have set a large sampling range for $q_0$, resulting in a lower mean value for $H_0$. At the same time, $j_0$ also exhibits degeneracy with $H_0$ in a similar manner. Therefore, we attempt to jointly constrain the four parameters $\{{H}_0,q_0,j_0,\Omega_{\rm{b,0}}\}$ using the FRB data in conjunction with Pantheon dataset.

The blue contour in Fig. \ref{fig:result1} represents the joint constraints from FRB+Pantheon datasets, where we set $f_{\rm{IGM,0}}=0.83$ and $\alpha_{\rm{IGM}}=0$ (case (i)). For the Hubble constant ${H}_0$, we obtain a constraint result of ${H}_0=65.5^{+6.4}_{-5.4}$ ${\rm{km~s^{-1}~Mpc^{-1}}}$ $(68\%$ ${\rm{C.L.}})$ and ${H}_0=65^{+10}_{-10}$ ${\rm{km~s^{-1}~Mpc^{-1}}}$ $(95\%$ ${\rm{C.L.}})$. Comparing this result to the constraint obtained using only the Pantheon dataset, it is clear that this constraint is derived from the constraints imposed by FRBs. The parameters $q_0=-0.50\pm 0.20$ and $j_0=-0.1^{+2.0}_{-2.5}$ are obtained. Compared to the constraint results from using only the Pantheon dataset, the addition of FRBs data results in similar accuracy. This may be due to the limited number of currently available localized FRBs, all within the redshift range of $z\leqslant0.66$.

As more FRBs with higher redshifts are observed in the future, it will be necessary to further investigate the accuracy of different forms of the cosmographic approach and the order of expansion, as discussed in Sec. \ref{sec:cosmography}. This is related to the number of predicted FRBs and their redshift distribution function, but a detailed discussion of these topics is beyond the scope of this paper.

\subsection{The Influence of Model Selection and Parameter Degeneracy}\label{sec:result2}

The model selection for constructing the likelihood of FRBs clearly affects the estimation of cosmological parameters. In this paper, we use the results from the IllustrisTNG simulation to obtain $C_0$ and $\sigma_{\rm{IGM}}$ for ${\rm{DM}}_{\rm{IGM}}$. Alternatively, assuming $F$ as a constant is also a possibility. \citet{Baptista:2023uqu} constrains ${\rm{log}}_{10}F$ from 78 FRBs (21 with redshifts), resulting in a value of ${\rm{log}}_{10}F=-0.48^{+0.26}_{-0.18}$. This consistency with the IllustrisTNG simulation results is shown within the redshift range from 0.4 to 2. However, at lower redshifts, the constrained $F$ value from FRB data is a smaller mean value than that from the simulations. For ${\rm{DM}}_{\rm{host}}$, this study relies on the results from IllustrisTNG simulation. Alternatively, \citet{Macquart:2020lln} treat $e^{\mu}$ and $\sigma_{\rm{host}}$ as constants, which can be determined through data constraints. During the writing of this article, \citet{Fortunato:2023deh} utilized 23 FRBs to constrain cosmographic parameters and achieved higher precision constraints using two different model selection of FRB compared to our study. In this study, we solely considered a specific set of model combinations discussed in Sec. \ref{sec:likelihood} without further analysis of additional combinations. If future observations yield models inconsistent with those in this paper or propose more theoretically general models, further analysis will be necessary to assess the impact of different $\rm{DM}$ models on constraining cosmological parameters.

\begin{figure}
	\includegraphics[scale=0.4]{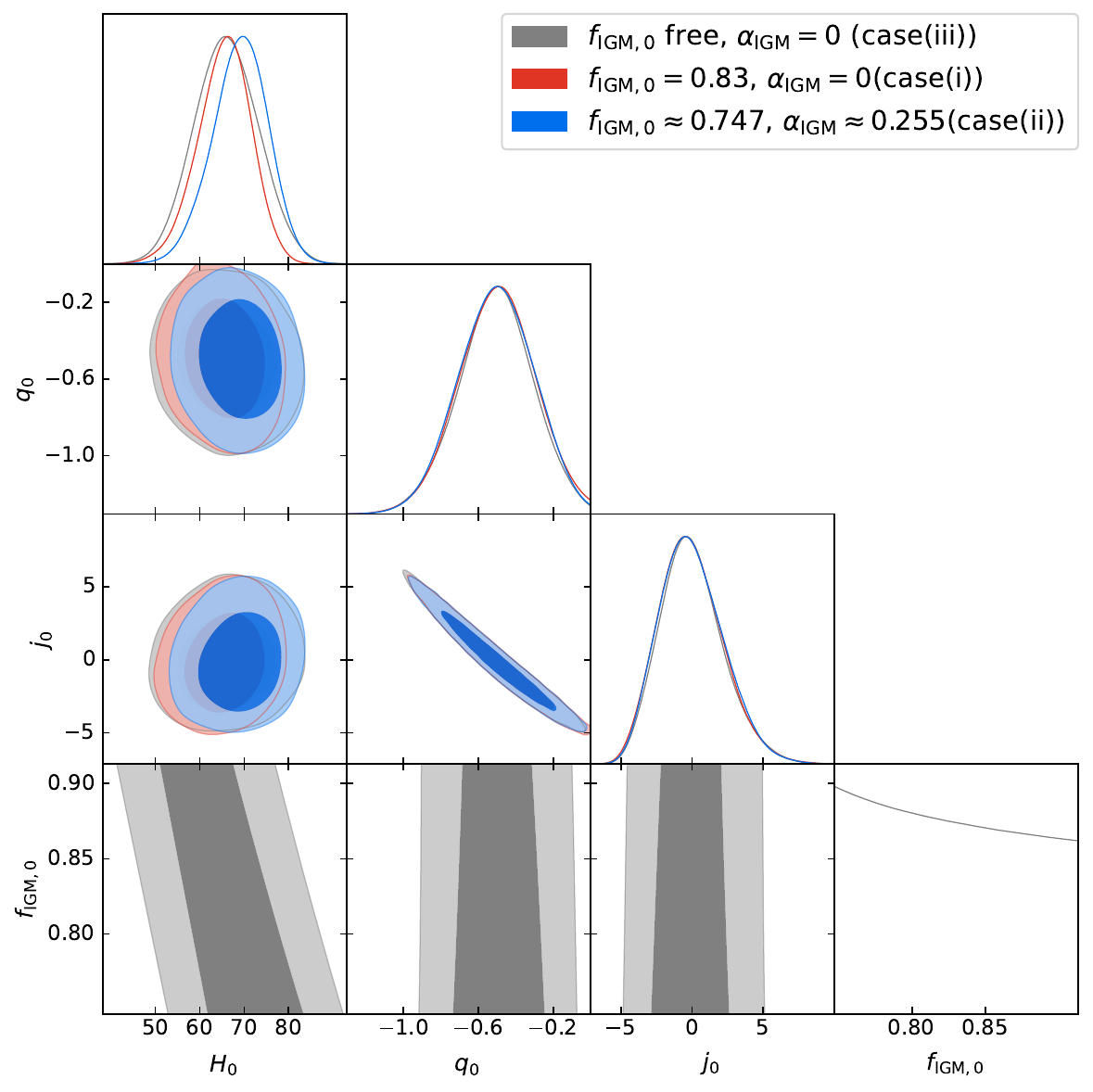}
	\caption{The contour plot shows the results of the 68\% and 95\% confidence intervals constrained by the same Pantheon+FRB data. To investigate the impact of the prior parameter $f_{\rm{IGM}}(z)$ on the posterior distribution of $H_0$, we set $\Omega_{\rm{b,0}}$ as a uniform prior with a sampling range of $1\sigma$, and consider three different scenarios for $f_{\rm{IGM}}(z)$.}
	\label{fig:result2}
\end{figure}

On the basis of the fixed model selection in this paper, an obvious issue is the parameter degeneracy between cosmological parameters and cosmographic parameters, which has been discussed in Sec. \ref{sec:result1} for $H_0$, $q_0$, and $j_0$. Additionally, we need to further discuss the influence of cosmological parameters $f_{\rm{IGM}}$ and $\Omega_{\rm{b,0}}$ on $H_0$. For $f_{\rm{IGM}}$, we consider three possible evolutionary forms and sample their posterior distributions. In this case, we treat $\Omega_{\rm{b,0}}$ as a free parameter and sample its $1\sigma$ range. The posterior distributions of the parameters are presented in Tab. \ref{tab:2}, while Fig. \ref{fig:result2} displays the contour plot of the posterior distributions obtained from the chains. Here, the primary focus is on discussing the impact of $f_{\rm{IGM}}$ on the posterior distribution of $H_0$, therefore the posterior distribution of $\Omega_{\rm{b,0}}$ is not displayed in the contour.

\begin{table}
	\centering
	\caption{The posterior distributions are constrained by the same Pantheon+FRB data. To investigate the impact of the prior on parameter $f_{\rm{IGM}}(z)$ on the posterior distribution of $H_0$, we set $\Omega_{\rm{b,0}}$ as a uniform prior with a sampling range of $1\sigma$. Additionally, $f_{\rm{IGM}}(z)$ is considered under three different scenarios, the case (i) represents $f_{\rm{IGM,0}}=0.83$ and $\alpha_{\rm{IGM}}=0$, the case (ii) represents $f_{\rm{IGM,0}}\approx0.747$ and $\alpha_{\rm{IGM}}(\alpha)\approx0.255$, the case (iii) represents $f_{\rm{IGM,0}}$ as a free parameter and $\alpha_{\rm{IGM}}=0$..}
	\begin{tabular}{l|c|c|c|c} 
		\hline
		Prior on $f_{\rm{IGM},0}$ & ${\rm{H}}_0$& $q_0$ & $j_0$  & $f_{\rm{IGM},0}$ \\
		\hline
		case (i) & $65.5^{+6.4}_{-5.4}        $ & $-0.50\pm 0.20             $& $-0.1^{+2.0}_{-2.5}        $ &\\
		case (ii) & $69.0^{+6.7}_{-5.7}        $ & $-0.51\pm 0.20             $ & $0.0^{+1.9}_{-2.5}         $   &     \\
		case (iii) & $66.0\pm 7.3               $& $-0.50\pm 0.20             $ & $0.0^{+1.9}_{-2.4}         $   &  ---  \\
		\hline
	\end{tabular}
	\label{tab:2}
\end{table}

As shown in Sec. \ref{sec:result1}, the main constraints on $q_0$ and $j_0$ come from the Pantheon dataset. The three different scenarios of $f_{\rm{IGM}}$ have a minimal impact on the posterior distributions of $q_0$ and $j_0$. When comparing the results of case (i) and case (ii), we observe that the evolving $f_{\rm{IGM}}$ we implemented results in a notably larger mean value for $H_0$. For the 18 FRB data utilized in this paper, neglecting the effects of integrating over redshift allows us to calculate the evolution of $f_{\rm{IGM}}$ from 0.747 to 0.848 as redshift from 0 to 0.66. Although the ${\rm{DM}}_{\rm{IGM}}$ model employed in this paper exhibits a smaller $\sigma_{\rm{IGM}}$ at higher redshifts, there is still a significant number of FRBs concentrated at lower redshifts with small $f_{\rm{IGM}}$ values. Additionally, Eq. \ref{eq:dmigm} and the posterior distribution from case (iii) in Fig. \ref{fig:result2} both reveal a negative correlation between $H_0$ and $f_{\rm{IGM}}$ in our parameter formulation, leading to a substantially larger $H_0$ in case (ii). When comparing case (i) and case (iii), we observe that introducing a free parameter $f_{\rm{IGM,0}}$ noticeably increases the $1\sigma$ error on $H_0$. Furthermore, even though the sampling range of $f_{\rm{IGM,0}}$ as a free parameter in case (iii) encompasses possible values of evolving $f_{\rm{IGM}}$, the prior range for $f_{\rm{IGM,0}}$ still results in a considerable deviation in the mean value of the posterior distribution. Consequently, this suggests that assuming a constant value and specifying a prior interval for $f_{\rm{IGM}}$, when it actually evolves with redshift, may lead to significant deviations in the posterior estimates of cosmological parameters. This emphasizes the necessity of delving into a deeper understanding of the potential evolution of $f_{\rm{IGM}}$ with redshift.

\begin{figure}
	\includegraphics[scale=0.4]{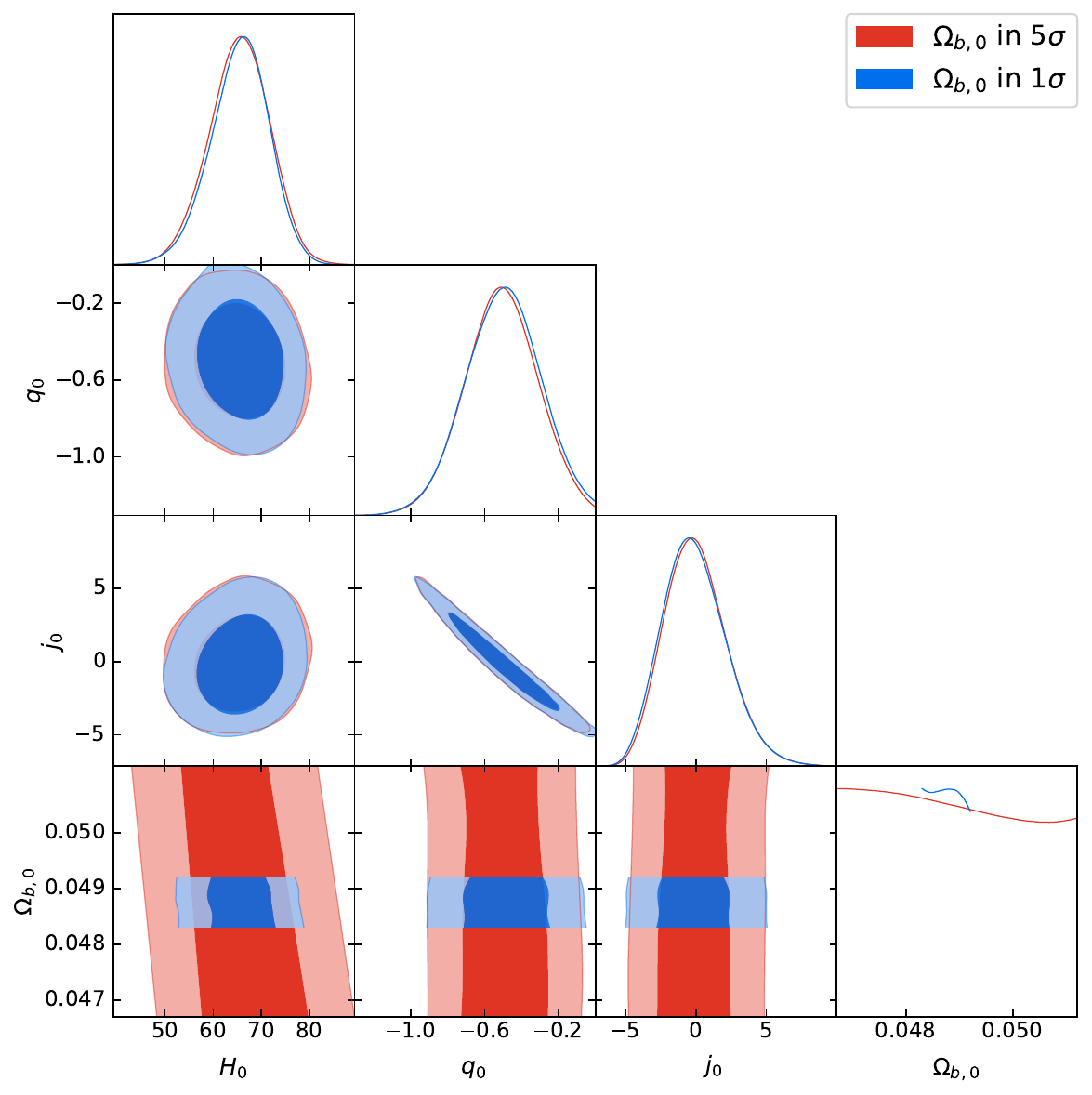}
	\caption{The contour plot illustrates the 68\% and 95\% confidence intervals obtained from the same Pantheon+FRB data. To explore the impact of the prior parameter $\Omega_{\rm{b,0}}$ on the posterior distribution of H0, we set $f_{\rm{IGM,0}}(z)$ to be in case (i) and provide sampling ranges of $1 \sigma$ and $5 \sigma$ for $\Omega_{\rm{b,0}}$, respectively.}
	\label{fig:result3}
\end{figure}

$\Omega_{\rm{b,0}}$ also has a parameter degeneracy with $H_0$. Due to the insufficient sample size of FRBs, the constraints on $\Omega_{\rm{b,0}}$ are limited in this paper. Therefore, we investigate the influence of $\Omega_{\rm{b,0}}$ on $H_0$ through two factors: the prior range for sampling $\Omega_{\rm{b,0}}$ and the selection of the prior interval for $\Omega_{\rm{b,0}}$. The former factor affects the estimation error of cosmological parameters, while the latter impacts the mean value of the parameters. To evaluate the effect of the sampling range on the posterior distribution of $H_0$, we set $f_{\rm{IGM,0}}=0.83$ (case (i)) and sample $\Omega_{\rm{b,0}}$ within the $1\sigma$ and $5\sigma$ prior intervals. The posterior distributions of the parameters are presented in Tab. \ref{tab:3}, and Fig. \ref{fig:result3} shows the corresponding contours.

\begin{table}
	\centering
	\caption{The posterior distribution is constrained by the same Pantheon+FRB data. To explore the impact of the prior parameter $\Omega_{\rm{b,0}}$ on the posterior distribution of $H_0$, we set $f_{\rm{IGM,0}}(z)$ to be in case (i), and provide sampling ranges of $1 \sigma$ and $5 \sigma$ for $\Omega_{\rm{b,0}}$, respectively.}
	\begin{tabular}{l|c|c|c|c} 
		\hline
		Prior on $\Omega_{\rm{b,0}}$ & ${\rm{H}}_0$& $q_0$ & $j_0$ & $\Omega_{\rm{b,0}}$  \\
		\hline
		$(0.0483,0.0492)$& $65.5^{+6.4}_{-5.4}        $ & $-0.50\pm 0.20             $& $-0.1^{+2.0}_{-2.5}        $& --- \\	
		$(0.0467,0.0513)$ & $65.4^{+6.6}_{-5.9}        $ & $-0.51\pm 0.20             $ & $0.0^{+1.9}_{-2.4}         $ & ---    \\		
		\hline
	\end{tabular}
	\label{tab:3}
\end{table}

The content depicted in Fig, \ref{fig:result3} aligns with the scenario presented in Eq. \ref{eq:dmigm}, where $\Omega_{\rm{b,0}}$ and $H_0$ exhibit a negative correlation in our model. However, due to the current limitations of FRB data in simultaneously constraining $H_0$ and $\Omega_{\rm{b,0}}$, a broader prior range for $\Omega_{\rm{b,0}}$ leads to increased uncertainty in $H_0$. Nonetheless, this can be subjected to quantitative analysis. Presently, expanding the prior interval of $\Omega_{\rm{b,0}}$ from $1\sigma$ to $5\sigma$ results in a 6$\%$ increase in the $1\sigma$ error of $H_0$, although no significant impact is observed. Regarding the influence of prior interval selection on the mean value of $H_0$, it is impossible to eliminate this effect within the scope of this study since FRB data alone can not adequately constrain both $H_0$ and $\Omega_{\rm{b,0}}$. It is crucial to acknowledge that the posterior distribution of $H_0$ in this study is contingent upon the assigned prior range of $\Omega_{\rm{b,0}}$. Additional FRB data or alternative observations are needed to provide prior information on $\Omega_{\rm{b,0}}$ in order to resolve this degeneracy.

\subsection{Discussion and Forecast}

Compared to our study presented in this paper, ${H}_0=73^{+12}_{-8}$ ${\rm{km~s^{-1}~Mpc^{-1}}}$ is constrained by 16 localized FRBs and 60 unlocalized FRBs (\cite{James:2022dcx}), Ref. \citet{Zhao:2022yiv} used 6 unlocalized ASKAP FRB data to constrain $H_0 = 71.7^{+8.8}_{-7.4}$ ${\rm{km~s^{-1}~Mpc^{-1}}}$ in the simulation-based case and $H_0 = 71.5^{+10.0}_{-8.1}$ ${\rm{km~s^{-1}~Mpc^{-1}}}$ in the observation-based case, while ${H}_0=62.3\pm9.1$ ${\rm{km~s^{-1}~Mpc^{-1}}}$ is constrained by 9 localized FRBs (\cite{Hagstotz:2021jzu}). It is worth noting that \citet{Liu:2022bmn} and \citet{wu20228} also use 18 FRB data to constrain ${H}_0=71\pm3$ ${\rm{km~s^{-1}~Mpc^{-1}}}$ and ${H}_0=68.81^{+4.99}_{-4.33}$ ${\rm{km~s^{-1}~Mpc^{-1}}}$, respectively. \citet{Liu:2022bmn} employs a model-independent method to constrain $H_0$ from FRBs in combination with 19 cosmic chronometers (CC) data, while \citet{wu20228} constrain $H_0$ based $\Lambda$CDM model, and treating $\Omega_m$ and $\Omega_bh^2$ as uniform prior distribution with the $1\sigma$ error range prior given by CMB for sampling to obtain the posterior distribution. In contrast, our approach in this paper involves using the cosmographic approach to expand the series at $z=0$, and we use the uncalibrated SNe Ia Pantheon dataset in this paper to provide priors for higher-order parameters $q_0$ and $j_0$.

Regarding the precision of the constraints in this paper, our joint constraints using the Pantheon dataset with marginalization of $M^{\prime}-\mu_0$ only yield low precision for $q_0$ and $j_0$, resulting in larger errors in $H_0$. However, this approach avoids relying on the prior of $H_0$ given by additional data. In terms of cosmological model dependence, our study is dependent on $\Lambda$CDM model, despite the cosmographic approach being a model-independent approximation method. Our result rely on the uniform posterior range of $\Omega_{\rm{b,0}}$ sampling, and for the model selection of FRB used in this paper, where $\rm{DM}_{\rm{IGM}}$ and $\rm{DM}_{\rm{host}}$ are obtained from IllustrisTNG simulations, which are based on $\Lambda$CDM model. However, as discussed in Sec. \ref{sec:result2}, if there are more observationally consistent model in the future, the approach presented in this paper could be completely independent of cosmological models. On the other hand, our results show that the obtained $H_0=65.5^{+6.4}_{-5.4}$ ${\rm{km~s^{-1}~Mpc^{-1}}}$ is consistent within the $1\sigma$ range of the IllustrisTNG simulation, indicating no model conflict.

When comparing our result with the ${H}_0 = 67.4\pm0.5 $ ${\rm{km~s^{-1}~Mpc^{-1}}}$ obtained by constraining the $\Lambda$CDM model using CMB and the ${H}_0 = 73.04\pm 1.04$ ${\rm{km~s^{-1}~Mpc^{-1}}}$ given by the SH0ES project using low-redshift Cepheid variables, it can be observed that the ${H}_0=65.5^{+6.4}_{-5.4} $ ${\rm{km~s^{-1}~Mpc^{-1}}}$ obtained by constraining the cosmographic approach using FRBs data is biased towards a smaller result that is closer to the $\Lambda$CDM mocel constraints derived from CMB or BAO. If we consider the case (ii) scenario for $f_{\rm{IGM}}(z)$, the posterior distribution of $H_0=69.0^{+6.7}_{-5.7}$ ${\rm{km~s^{-1}~Mpc^{-1}}}$ tends towards a middle value, and the mean value becomes closer to the results obtained from SNe Ia calibrated using TRGB.

\begin{figure}
	\includegraphics[scale=0.5]{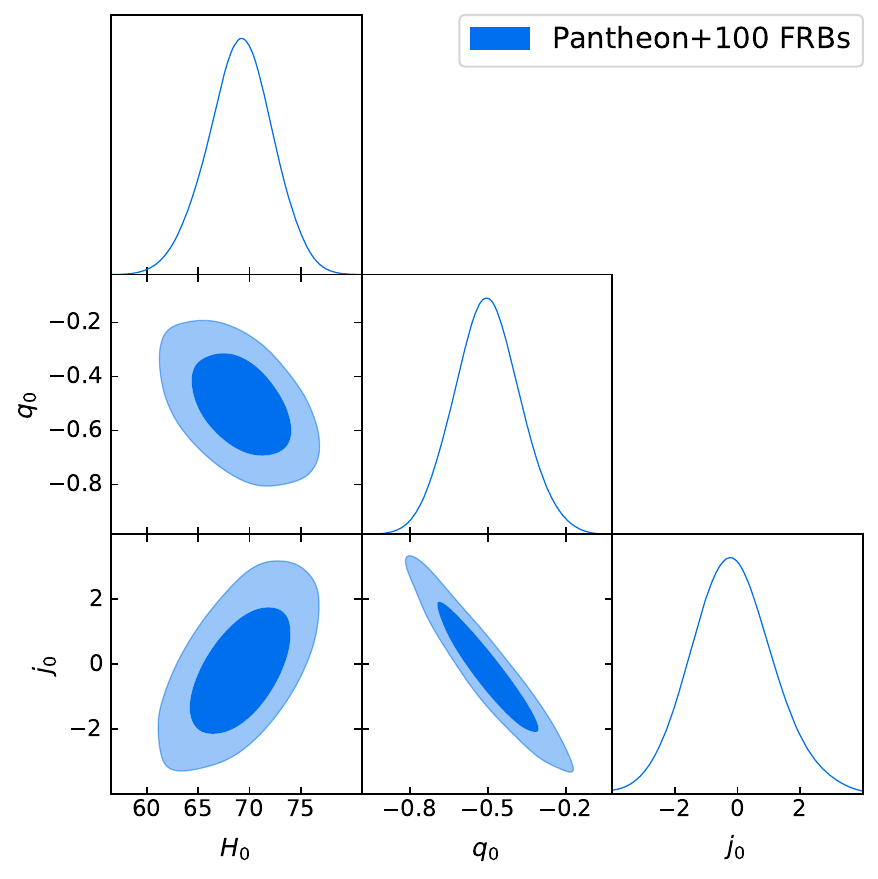}
	\caption{Based on the model in this paper, the 68\% and 95\% confidence intervals were obtained by combining 100 mock FRB data with the Pantheon dataset. The results show a constraint accuracy of 4.6\% for $H_0$, demonstrating the conservative estimate of the $H_0$ constraint precision achievable through the method presented in this study for future FRB observations.}
	\label{fig:result4}
\end{figure}

Furthermore, we also used 100 mock FRBs to estimate the precision of constraints on $H_0$ that can be achieved in the near future using the method proposed in this paper. We utilized the model presented in Sec. \ref{sec:likelihood}, simulating ${\rm{DM}}_{\rm{IGM}}$ and ${\rm{DM}}_{\rm{host}}$ separately. For the cosmological parameters, we considered case (i) for $f_{\rm{IGM}}(z)$ and $1\sigma$ range for $\Omega_{\rm{b,0}}$, with the baseline value of $H_0=70$ ${\rm{km~s^{-1}~Mpc^{-1}}}$. Since the constraint results derived from the mock data are solely intended to illustrate the precision issue, we adopted the posterior distribution of $q_0$ and $j_0$ from the Pantheon as the baseline. The ratio of the three types of FRBs in ${\rm{DM}}_{\rm{host}}$ was established as $1:1:3$, closely resembling the current ratio of localized FRB data. To accelerate the sampling speed, we set the sampling range of cosmographic parameters for the mock data as follows, ${H}_0\in \mathcal{U}(55,85)$, $q_0\in \mathcal{U}(-1,0)$, $j_0\in \mathcal{U}(-4,4)$ . Fig. \ref{fig:result4} displays the constraint results obtained from the combination of 100 mock FRBs and Pantheon. We observed that $H_0=69.1\pm 3.2$ ${\rm{km~s^{-1}~Mpc^{-1}}}$, $q_0=-0.50\pm 0.13$ and  $j_0=-0.2\pm 1.3$, achieving a precision of $4.6\%$ for $H_0$. In contrast to constraints obtained solely from Pantheon, the inclusion of 100 mock data considerably enhances the precision of higher-order cosmographic parameters $q_0$ and $j_0$.

\section{Conclusions}\label{sec:conclusions}

Since Steven Weinberg first proposed the cosmographic approach, it has been used to construct various mathematical expansions of physical quantities, such as the comoving angular diameter distance related to BAO, the luminosity distance related to Gamma ray bursts (GRBs) and SNe Ia, the Hubble parameter related to CC data. In this paper, we use the cosmographic approach to obtain a new form of $\langle{\rm{DM}}_{\rm{IGM}}(z)\rangle$ related to FRBs. We analyze the properties of this form and compare it to the standard $\Lambda$CDM model, and then validate its feasibility using data from 18 localized FRBs combined with the Pantheon dataset.

In order to obtain the posterior distribution of the parameters, we utilize the model described in Sec. \ref{sec:likelihood}. We set case (i) for $f_{\rm{IGM}}(z)$, and a uniform priors with $1\sigma$ range for $\Omega_{\rm{b,0}}$, to obtain the posterior distribution of the parameters. We combine 18 localized FRBs with the Pantheon dataset to constrain cosmographic parameters and obtain ${H}_0=65.5^{+6.4}_{-5.4}$ ${\rm{km~s^{-1}~Mpc^{-1}}}$ $(68\%$ ${\rm{C.L.}})$, $q_0=-0.50\pm 0.20$ and $j_0=-0.1^{+2.0}_{-2.5}$, whereby $q0$ and $j0$ are primarily constrained by Pantheon. The results obtained from FRB alone reveal the degeneracy among cosmographic parameters, indicating a negative correlation between $H_0$ and $q_0$, $j_0$. Additionally, the choice of FRB models and the degeneracy between cosmological parameters also influence the results of our study. Determining and discussing the optimal model selection may require more astronomical observations or simulations, which we do not extensively elaborate on in this paper. We specifically employ one model selection from Sec. \ref{sec:likelihood} to explore the impact of $f_{\rm{IGM}}(z)$ and $\Omega_{\rm{b,0}}$ priors on the posterior distribution of $H_0$. We observe that enlarging the prior range for $f_{\rm{IGM,0}}$ and $\Omega_{\rm{b,0}}$ does not significantly increase the $1\sigma$ error of the posterior distribution of $H_0$. However, in cases where FRB can not simultaneously constrain $H_0$, $f_{\rm{IGM,0}}$ and $\Omega_{\rm{b,0}}$, different prior values for $f_{\rm{IGM,0}}$ and $\Omega_{\rm{b,0}}$ yield distinct posterior distributions of $H_0$ due to parameter degeneracy. The outcomes of this study rely on the prior assumptions of these two parameters as well.

Using FRB data to constrain cosmographic parameters also poses several challenges. As we discovered in Sec. \ref{sec:cosmography} of our analysis, constraining cosmographic parameters with FRBs depends on the number of FRBs and their redshift distribution. This necessitates a thorough examination of various expansion bases. Furthermore, more precise measurements of parameters such as $\Omega_{\rm{b,0}}$ and $f_{\rm{IGM}}(z)$ can enhance the accuracy of FRB constraints on $H_0$ and other cosmographic parameters. This will require more observation data and resolving parameter degeneracies via joint analysis.

\section*{Acknowledgements}
This work is supported in part by National Natural Science Foundation of China under Grant No. 12075042 and No. 11675032, Jiangsu Funding Program for Excellent Postdoctoral Talent (20220ZB59) and Project funded by China Postdoctoral Science Foundation (2022M721561). We thank G. Q. Zhang, Yang Liu and Ze-Wei Zhao for discussion.

\bibliographystyle{mnras}
\bibliography{manuscript} 

\appendix

\bsp	
\label{lastpage}
\end{document}